\documentclass[preprint,12pt,authoryear]{elsarticle}
\usepackage[utf8]{inputenc}
\usepackage{amssymb}
\usepackage{amsmath}
\usepackage{graphicx}
\usepackage{amsmath}
\usepackage{amssymb}
\usepackage{booktabs}

\usepackage{amsfonts}
\usepackage{float}
\usepackage{array}
\usepackage{color}
\usepackage{tabularx}
\usepackage{longtable}
\usepackage{tabu}
\usepackage{multirow}
\usepackage{url}
\usepackage{bm}
\usepackage[noend]{algpseudocode}
\usepackage{algorithmicx,algorithm}
\usepackage{adjustbox}
\usepackage{makecell}
\usepackage{diagbox}
\definecolor{myblue}{rgb}{0.21,0.49,0.74}
\journal{Expert Systems with Applications}

\begin{document}

\begin{frontmatter}

% \biboptions{authoryear,round}
% \bibliographystyle{model5-names}

%% Title, authors and addresses

%% use the tnoteref command within \title for footnotes;
%% use the tnotetext command for theassociated footnote;
%% use the fnref command within \author or \affiliation for footnotes;
%% use the fntext command for theassociated footnote;
%% use the corref command within \author for corresponding author footnotes;
%% use the cortext command for theassociated footnote;
%% use the ead command for the email address,
%% and the form \ead[url] for the home page:
%% \title{Title\tnoteref{label1}}
%% \tnotetext[label1]{}
%% \author{Name\corref{cor1}\fnref{label2}}
%% \ead{email address}
%% \ead[url]{home page}
%% \fntext[label2]{}
%% \cortext[cor1]{}
%% \affiliation{organization={},
%%             addressline={},
%%             city={},
%%             postcode={},
%%             state={},
%%             country={}}
%% \fntext[label3]{}

\title{Enhancing Blind Video Quality Assessment with Rich Quality-aware Features}

%% use optional labels to link authors explicitly to addresses:
%% \author[label1,label2]{}
%% \affiliation[label1]{organization={},
%%             addressline={},
%%             city={},
%%             postcode={},
%%             state={},
%%             country={}}
%%
%% \affiliation[label2]{organization={},
%%             addressline={},
%%             city={},
%%             postcode={},
%%             state={},
%%             country={}}

\author[ecnu]{Wei Sun\corref{cor1}}
\ead{wsun@cee.ecnu.edu.cn}

\author[sjtu]{Linhan Cao}
\ead{caolinhan@sjtu.edu.cn}

\author[sjtu]{Jun Jia}
\ead{jiajun0302@sjtu.edu.cn}

\author[sjtu]{Zhichao Zhang}
\ead{liquortect@sjtu.edu.cn}

\author[sjtu]{Zicheng Zhang}
\ead{zzc1998@sjtu.edu.cn}

\author[sjtu]{Xiongkuo Min}
\ead{minxiongkuo@sjtu.edu.cn}

\author[sjtu]{Guangtao Zhai}
\ead{zhaiguangtao@sjtu.edu.cn}

\affiliation[ecnu]{organization={East China Normal University},
            city={Shanghai},
            country={China}}

\affiliation[sjtu]{organization={Shanghai Jiao Tong University},
            city={Shanghai},
            country={China}}

\cortext[cor1]{Corresponding author. Email: wsun@cee.ecnu.edu.cn}

%% Abstract
\begin{abstract}
Blind video quality assessment (BVQA) is a highly challenging task due to the intrinsic complexity of video content and visual distortions, especially given the high popularity of social media videos, which originate from a wide range of sources, and are often processed by various compression and enhancement algorithms. While recent BVQA and blind image quality assessment (BIQA) studies have made remarkable progress, their models typically perform well on the datasets they were trained on but generalize poorly to unseen videos, making them less effective for accurately evaluating the perceptual quality of diverse social media videos. In this paper, we propose \textbf{R}ich \textbf{Q}uality-aware features enabled \textbf{V}ideo \textbf{Q}uality \textbf{A}ssessment \textbf{(RQ-VQA)}, a simple yet effective method to enhance BVQA by leveraging rich quality-aware features extracted from off-the-shelf BIQA and BVQA models. Our approach exploits the expertise of existing quality assessment models within their trained domains to improve generalization. Specifically, we design a multi-source feature framework that integrates:
(1) \textbf{Learnable spatial features} from a base model fine-tuned on the target VQA dataset to capture domain-specific quality cues;
(2) \textbf{Temporal motion features} from the fast pathway of SlowFast pre-trained on action recognition datasets to model motion-related distortions;
(3) \textbf{Spatial quality-aware features} from BIQA models trained on diverse IQA datasets to enhance frame-level distortion representation; and
(4) \textbf{Spatiotemporal quality-aware features} from a BVQA model trained on large-scale VQA datasets to jointly encode spatial structure and temporal dynamics. These features are concatenated and fed into a multi-layer perceptron (MLP) to regress them into quality scores. Experimental results demonstrate that our model achieves state-of-the-art performance on three public social media VQA datasets. Our method is extremely simple in both architecture and training, and can be easily extended to other quality assessment tasks with different quality-aware feature sources. Moreover, \textbf{RQ-VQA won first place in the CVPR NTIRE 2024 Short-form UGC Video Quality Assessment Challenge}. The code is available at \url{https://github.com/sunwei925/RQ-VQA.git}.

\end{abstract}

%%Graphical abstract
% \begin{graphicalabstract}
% %\includegraphics{grabs}
% \end{graphicalabstract}

%%Research highlights
% \begin{highlights}
% \item We propose RQ-VQA, a simple and effective BVQA framework for social media videos.
% \item RQ-VQA integrates spatial, temporal, and spatiotemporal quality cues modularly.
% \item It achieves strong generalization without requiring large-scale task-specific data.
% \item RQ-VQA attains state-of-the-art results on multiple social media VQA benchmarks.
% \item Our method won first place in the CVPR NTIRE 2024 UGC Video Quality Challenge.
% \end{highlights}

%% Keywords
\begin{keyword}
Blind Video Quality Assessment, Neural Network, Quality-aware Features, Feature Fusion, Social Media Video.
%% keywords here, in the form: keyword \sep keyword

%% PACS codes here, in the form: \PACS code \sep code

%% MSC codes here, in the form: \MSC code \sep code
%% or \MSC[2008] code \sep code (2000 is the default)

\end{keyword}

\end{frontmatter}

%% Add \usepackage{lineno} before \begin{document} and uncomment 
%% following line to enable line numbers
%% \linenumbers

%% main text
%%

%% Use \section commands to start a section
\section{Introduction}
\label{sec:intro}
Blind video quality assessment (BVQA)~\citep{min2024perceptual} aims to predict the perceptual quality of a video without access to any reference information (\textit{i.e.,} high-quality source videos). With the rapid proliferation of social media platforms and the surge in user-generated content (UGC), BVQA has become increasingly critical in video processing systems for streaming media applications, ensuring that end-users experience high-quality videos and superior Quality of Experience (QoE). Towards this goal, numerous BVQA models have been developed to achieve stronger correlations with human subjective opinions, including knowledge-driven approaches~\citep{saad2014blind, korhonen2019two, tu2021ugc, tu2021rapique} and and data-driven approaches~\citep{li2019quality,yi2021attention,sun2021deep,ying2021patch,li2022blindly,wu2022fast,liu2023ada,sun2022deep, sun2024analysis}.

Although knowledge-driven BVQA models~\citep{saad2014blind, korhonen2019two, tu2021ugc, tu2021rapique, ebenezer2021chipqa} offer better interpretability by incorporating hand-crafted features and perceptual principles inspired by the human visual system, they often suffer from relatively low prediction accuracy and higher computational overhead compared to data-driven approaches. This performance gap largely stems from the inherent complexity of human visual perception, which is difficult to fully capture through manually designed features and heuristic rules. With the rapid advances in deep neural networks (DNNs), data-driven BVQA models have emerged as a dominant paradigm, learning powerful quality-aware feature representations directly from large-scale annotated datasets. These models have demonstrated state-of-the-art performance across different video domains, including professionally generated content (PGC) videos with synthetic distortions~\citep{lu2023bh}  and UGC videos with realistic distortions~\citep{sun2022deep,wu2022fast,wang2021rich,liu2023ada}.

The success of data-driven BVQA models can be largely attributed to two key factors.
First, the adoption of increasingly advanced neural network architectures has significantly enhanced their capacity to model complex relationships between video content, distortion patterns, and perceived quality. These architectures include convolutional neural network (CNN)-based methods (e.g., VSFA~\citep{li2019quality}, SimpleVQA~\citep{sun2022deep}, Li22~\citep{li2022blindly}), which capture spatial and temporal features through hierarchical convolutional operations; Transformer-based methods (e.g., StarVQA~\citep{xing2022starvqa}, FAST-VQA~\citep{wu2022fast}), which leverage self-attention mechanisms to model long-range dependencies in both spatial and temporal dimensions; and more recently, large multimodal (LMM)-based approaches (e.g., Q-Align~\citep{wu2022fast}, LMM-VQA~\citep{ge2025lmm}), which incorporate cross-modal reasoning capabilities to better align visual quality cues with semantic context. Second, the growing availability of large-scale, subjectively annotated video quality assessment (VQA) datasets has provided rich supervision for training deep models. For instance, widely used UGC VQA datasets such as KonViD-1k~\citep{hosu2017konstanz}, YouTube-UGC~\citep{wang2019youtube}, and LSVQ~\citep{ying2021patch} contain thousands to tens of thousands of in-the-wild videos annotated with human-rated quality scores.

As data-driven methods, their performance is heavily dependent on the human-rated VQA datasets used for training. However, the videos in most mainstream VQA datasets~\citep{nuutinen2016cvd2014, ghadiyaram2017capture, hosu2017konstanz, sinno2018large, wang2019youtube, ying2021patch} were typically captured with outdated cameras or collected from video-sharing platforms several years ago. As a result, their distortion types and content often fail to reflect the characteristics of videos in modern streaming applications, particularly on social media, where shooting devices and video processing algorithms, including pre-processing, compression, and enhancement techniques, have advanced substantially. Consequently, BVQA models trained on these datasets tend to achieve good performance on their own training domains but generalize poorly to unseen videos, such as the millions of newly uploaded social media videos generated daily.

\begin{figure*}
    \centering
    \includegraphics[width=1\linewidth]{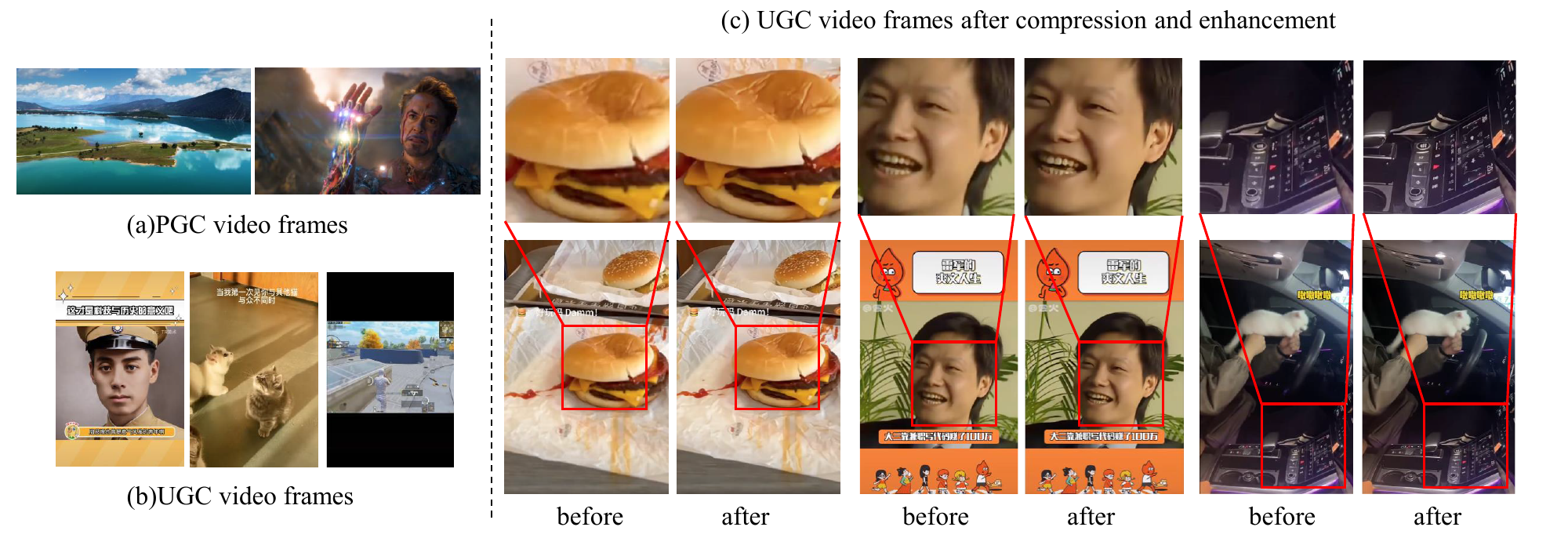}
    \caption{The comparison of PGC videos, UGC videos, and the processed UGC videos.}
    \label{fig:video_frames}
\end{figure*}

To address the challenges posed by the complex distortions and diverse content in social media videos, we propose Rich Quality-aware Video Quality Assessment (RQ-VQA), a simple yet effective blind VQA framework that leverages the complementary strengths of multiple pre-trained models. Instead of training a large-scale model from scratch, RQ-VQA extracts multi-source quality-aware features from both learnable and off-the-shelf networks to enhance generalization to unseen data. Specifically, the framework integrates (1) learnable spatial features from the target video domain and (2) frozen spatial, temporal, and spatiotemporal features from existing BIQA, BVQA, and action recognition models, capturing rich perceptual cues across multiple dimensions. These heterogeneous features are fused through a lightweight regression module to produce a final quality score. Owing to its modular design and simplicity, RQ-VQA achieves state-of-the-art performance on multiple public social-media VQA benchmarks and ranks first place in the CVPR NTIRE 2024 Short-form UGC Video Quality Assessment Challenge, demonstrating its robustness and extensibility.

The main contributions of this work can be summarized as follows:
\begin{itemize}
\item We introduce a multi-source feature integration framework for BVQA that effectively combines diverse quality-aware representations without relying on large-scale task-specific training.
\item We demonstrate that leveraging rich quality-aware features from existing pre-trained BIQA and BVQA models substantially improves generalization to real-world social-media videos.
\item We conduct comprehensive experiments and ablation studies on several social-media VQA datasets, validating the effectiveness, efficiency, and generalization capability of the proposed method.
\end{itemize}

\section{Related Work}
\label{sec:related_work}

\subsection{VQA Datasets}
Early VQA datasets primarily focus on synthetic distortions introduced by different video processing stages, such as spatiotemporal downsampling~\citep{li2019avc, mackin2015study, nasiri2015perceptual, madhusudana2021subjective, lee2021subjective}, compression~\citep{seshadrinathan2010study,de2010h,vu2014vis,li2019avc}, transmission~\citep{moorthy2012video,chen2014modeling,ghadiyaram2014study,duanmu2016quality}, etc. These datasets typically consist of a limited number of high-quality source videos and the corresponding distorted ones. Due to limited video content and not considering the realistic distortions, these datasets are not suitable for training general BVQA models. Therefore, recent VQA datasets~\citep{nuutinen2016cvd2014, ghadiyaram2017capture, hosu2017konstanz, sinno2018large, wang2019youtube, ying2021patch} have shifted focus towards realistic captured distortions. For example, LIVE-Qualcomm~\citep{ghadiyaram2017capture} consists of $208$ videos captured by $8$ smartphones across $54$ unique scenes. KoNViD-1k~\citep{hosu2017konstanz} consists of $1,200$ Internet videos selected from YFCC100M using a density-based fair sampling strategy over six attributes, with MOS ratings obtained via crowdsourcing. LIVE-VQC~\citep{sinno2018large} includes $585$ videos captured by $80$ mobile cameras, encompassing different lighting conditions and diverse levels of motion, each video corresponding to a unique scene. YouTube-UGC~\citep{wang2019youtube} contains $1,500$ $20$-second videos sampled from $1.5$ million YouTube videos across $15$ categories, using a density-based strategy over spatial, color, temporal, and chunk variation attributes. LSVQ~\citep{ying2021patch} consists of $38,811$ videos sampled from the Internet Archive and YFCC100M datasets by matching six video feature distributions. In general, these datasets have greatly promoted the development of objective BVQA models.

\begin{table*}
  \centering
  \renewcommand{\arraystretch}{1.25}
  \caption{Summary of eight VQA datasets (Duration in seconds). 
  ``In-the-wild'' denotes distortions introduced during video capture; 
  ``Compression'' denotes distortions caused by compression algorithms; 
  ``Enhancement'' denotes artifacts introduced by video enhancement algorithms; 
  ``In-laboratory'' means subjective experiments conducted in a controlled lab environment; 
  ``Crowdsourcing'' means subjective experiments conducted via online crowd platforms.}
  \label{overview_vqa_dataset}

  \resizebox{1\textwidth}{!}{
  \begin{tabular}{lcccccccc}
  \toprule[.15em]
    \multirow{1}{*}{Dataset} & \multirow{1}{*}{Year} & \multirow{1}{*}{\# of Videos} & \multirow{1}{*}{\# of Scenes} & \multirow{1}{*}{Resolution} & Duration & \multirow{1}{*}{Frame Rate} & \multirow{1}{*}{Distortion Types} & \multirow{1}{*}{Environment} \\
  \hline
    LIVE-Qualcomm \citep{ghadiyaram2017capture} & 2017 & 208 & 54 & 1080p & 15 & 30 & In-the-wild & In-laboratory \\
    KoNViD-1k \citep{hosu2017konstanz} & 2017 & 1,200 & 1,200 & 540p & 8 & 24, 25, 30 & In-the-wild & Crowdsourcing \\
    LIVE-VQC \citep{sinno2018large} & 2018 & 585 & 585 & 240p-1080p & 10 & 30 & In-the-wild & Crowdsourcing \\
    YouTube-UGC \citep{wang2019youtube} & 2019 & 1,500 & 1,500 & 360p-4K & 20 & 30 & In-the-wild & Crowdsourcing \\
    LSVQ \citep{ying2021patch} & 2021 & 38,811 & 38,811 & 99p-4K & 5-12 & $\leq$ 60 & In-the-wild & Crowdsourcing \\
    UGC-VIDEO~\citep{li2021user} & 2021 & 550 & 50 & 720p & 10 & 30 & In-the-wild, Compression & In-laboratory \\
    LIVE-WC~\citep{yu2021predicting} & 2021 & 275 & 55 & 360p-1080p & 10 & 30 & In-the-wild, Compression & In-laboratory \\
    TaoLive~\citep{zhang2023md} & 2023 & 3,344 & 418 & 720p, 1080p & 8 & 20 & In-the-wild, Compression & In-laboratory \\
    \multirow{2}{*}{KVQ~\citep{lu2024kvq}} & \multirow{2}{*}{2024} & \multirow{2}{*}{3,600} & \multirow{2}{*}{600} & \multirow{2}{*}{368p-4K} & \multirow{2}{*}{$\leq$ 12} & \multirow{2}{*}{$\leq$ 120} & In-the-wild & \multirow{2}{*}{In-laboratory} \\
    & & & & & & & Compression, Enhancement & \\
  \bottomrule[.15em]
  \end{tabular}
  }
\end{table*}

However, for videos on social media platforms such as Kwai and TikTok, visual quality is affected by both in-capture distortions and degradations introduced by complex video processing pipelines. To address these challenges, several recent works have constructed VQA datasets tailored to social media content. For example, Li \textit{et al.}\citep{li2021user} proposed the UGC-VIDEO dataset, which includes 50 TikTok source videos, each transcoded using H.264 and H.265 at five QP levels to simulate real-world compression. Yu \textit{et al.}\citep{yu2021predicting} developed the LIVE-WC dataset by downsampling 55 1080p videos from LIVE-VQC~\citep{sinno2018large} to four resolutions, followed by H.264 compression at 17 levels; 220 representative distorted videos were selected for subjective quality annotation. Zhang \textit{et al.}\citep{zhang2023md} introduced the TaoLive dataset, comprising 418 raw videos and 3,762 distorted versions generated using H.265 at eight CRF levels. More recently, Lu \textit{et al.}\citep{lu2024kvq} presented the KVQ dataset, which captures the full video processing workflow—including pre-processing, transcoding, and enhancement—using 600 user-uploaded social media videos and 3,600 processed counterparts. A detailed summary of these datasets is provided in Table~\ref{overview_vqa_dataset}.

In this paper, we focus on quality assessment for UGC videos that undergo multiple video processing steps—referred to as social media videos—which pose greater challenges for BVQA models due to diverse distortions introduced during both capture and post-processing.

%-------------------------------------------------------------------------
\subsection{BVQA Models}

As stated in Section \ref{sec:intro}, we can roughly divide the BVQA models into knowledge-driven methods and data-driven methods.

\noindent\textbf{Knowledge-driven BVQA models}~\citep{saad2014blind, mittal2015completely, korhonen2019two, tu2021ugc, tu2021rapique, ebenezer2021chipqa} utilize carefully designed handcrafted features to quantify the video quality. For example, V-BLIINDS \citep{saad2014blind} utilizes spatiotemporal natural scene statistics (NSS) models to quantify the NSS features of frame differences and motion coherency characteristics, and then regresses these features to video scores by support vector regressor (SVR). Mittal \textit{et al.} \citep{mittal2015completely} propose a training-free blind VQA model named VIIDEO that exploits intrinsic statistics regularities of natural videos to quantify disturbances introduced due to distortions. TLVQM \citep{korhonen2019two} extracts rich spatiotemporal features such as motion, jerkiness, blurriness, noise, blockiness, color, etc. from both high and low complexity levels. VIDEVAL \citep{tu2021ugc} employs the sequential forward floating selection strategy to choose a set of quality-aware features from typical BI/VQA methods, followed by training an SVR model to regress them into the video quality. TLVQM and VIDEVAL demonstrate that leveraging rich quality-aware handcrafted features enables the BVQA model to achieve better performance. In this paper, we show that combining diverse quality-aware features extracted from DNNs with a base DNN model can also achieve superior performance.

\vspace{0.2cm}

\begin{table*}[htbp]
\centering
\scriptsize
\renewcommand{\arraystretch}{1.05}
\caption{Summary of variables and notations used in the proposed RQ-VQA framework.}
\label{tab:notation}

\resizebox{\textwidth}{!}{
\begin{tabular}{ll}
\toprule
\textbf{Symbol} & \textbf{Definition} \\
\midrule
$\bm{x}= \{\bm x_i\}_{i=0}^{N-1}$ & Input video sequence \\
$\bm{x}_i$ & The $i$-th video frame \\
$H, W$ & Height and width of each frame \\
$N$ & Total number of frames in the video \\
$r$ & Frame rate (frames per second) \\

$\bm{z}=\{\bm{z}_i\}_{i=0}^{N_z-1}$ & Sampled key frames (1 frame per second) \\
$\bm{z}_i$ & The $i$-th key frame \\
$N_z$ & Number of sampled key frames \\

$\mathcal{V}=\{\bm{v}^{(i)}\}_{i=0}^{N_z-1}$ & Set of video chunks \\
$\bm{v}^{(i)}$ & The $i$$^{\text{th}}$ video chunk (contains $r$ consecutive frames) \\

$\bm{x}^f$ & Spatiotemporal fragments extracted via GMS sampling \\
$G_f$ & Number of grids per frame in GMS ($G_f \times G_f$ grids) \\
$\bm{t}_{\mathrm{LIQE}}$ & Textual prompt template used in LIQE \\
$\bm{t}_{\mathrm{Q\text{-}Align}}$ & Textual question–answer prompt used in Q-Align \\

$\mathcal{F}_i^{s}$ & Spatial features of the $i$-th key frame from the base model \\
$\mathcal{F}_i^{\mathrm{LIQE}}$ & Spatial quality-aware features of the $i$-th key frame from LIQE \\
$\mathcal{F}_i^{\mathrm{Q\text{-}Align}}$ & Spatial quality-aware features of the $i$-th key frame from Q-Align \\
$\mathcal{F}^{\mathrm{FAST\textrm{-}VQA}}$ & Spatiotemporal features from FAST-VQA \\
$\mathcal{F}_i^{t}$ & Temporal motion features of the $i$-th chunk from SlowFast \\
$\mathcal{F}_i$ & Final fused multi-source feature vector \\

$\hat{q}_i$ & Predicted local quality score for frame/chunk $i$ \\
$\hat{q}$ & Final predicted video quality score \\
$q$ & Ground-truth MOS score \\
$\mathcal{L}$ & PLCC-based training loss function \\
\midrule
MHSA($\cdot$) & Multi-head self-attention module \\
SwinB($\cdot$) & Swin Transformer-B feature extractor (w/o classifier) \\
LIQE($\cdot$) & LIQE quality feature extractor \\
Q-Align($\cdot$) & Q-Align hidden-layer feature extractor \\
FAST-VQA($\cdot$) & FAST-VQA spatiotemporal feature extractor \\
SlowFast($\cdot$) & SlowFast fast-pathway backbone \\
GMS($\cdot$) & Grid Mini-Cube Sampling operator \\
Cat($\cdot$) & Feature concatenation operator \\
MLP($\cdot$) & Two-layer regression network \\
GP($\cdot$) & Global average pooling operator \\
$\langle\cdot,\cdot\rangle$ & Inner product \\
$\|\cdot\|_2$ & Euclidean norm \\
mean($\cdot$) & Mean operator \\
\bottomrule
\end{tabular}
}
\end{table*}

\noindent\textbf{Data-driven BVQA methods}~\citep{li2019quality,yi2021attention,ying2021patch,sun2019mc360iqa,li2022blindly,wu2022fast,wang2021rich,sun2025empirical,sun2022deep, sun2024analysis,sun2024assessing,sun2024dual,sun2025efficient,sun2025compressedvqa} mainly leverage DNNs to extract the quality-aware features.
VSFA~\citep{li2019quality} first extracts semantic features from a pre-trained CNN model, followed by utilizing a gated recurrent unit (GRU) network to capture the temporal relationship among the semantic features of video frames. 
Yi \textit{et al.}~\citep{yi2021attention} propose an attention mechanism based BVQA model, which employs a non-local operator to handle uneven spatial distortion problems.  
Ying \textit{et al.}~\citep{ying2021patch} introduce a local-to-global region-based BVQA model, combing the quality-aware features extracted from a BIQA pre-trained and spatiotemporal features from a pre-trained action recognition network.
Li \textit{et al.}~\citep{li2022blindly} also employ the IQA model pre-trained on multiple databases to extract quality-aware spatial features and the action recognition model to extract temporal features, subsequently utilizing a GRU network is used to regress spatial and temporal features into the quality scores. Sun \textit{et al.}~\citep{sun2022deep, sun2024analysis} propose SimpleVQA, a BVQA framework that consists of a trainable spatial feature extraction module and a pre-trained motion feature extraction model. In this paper, we adopt SimpleVQA as our base model.
Wu \textit{et al.}~\citep{wu2022fast} propose FAST-VQA, which samples spatio-temporal grid mini-cubes from original videos and trains a fragment attention network consisting of a Swin transformer and the gated relative position biases in an end-to-end manner. Wu~\textit{et al.}~\citep{wu2023exploring} further propose DOVER, which integrates FAST-VQA with an aesthetics quality assessment branch to evaluate video quality from both technique and aesthetics perspectives. With the popularity of large multi-modality models (LMMs), some LMM-based quality assessment models~\citep{wu2023q,cao2025vqathinker, ge2025lmm, cao2025breaking, huang2024visualcritic} have been proposed to evaluate the image/video quality by providing predefined text prompts to LMMs.

Recently, there have been efforts to integrate various types of DNN features to enhance BVQA performance and provide explainability. For example, Wang \textit{et al.} \citep{wang2021rich} propose a feature-rich BVQA model that assesses quality from three aspects including compression level, video content, and distortion type, with each aspect evaluated by a separate neural network. Liu \textit{et al.}~\citep{liu2023ada} extract seven types of features extracted by EfficientNet-b7~\citep{tan2019efficientnet}, ir-CSN-152~\citep{tran2019video}, CLIP~\citep{radford2021learning}, Swin Transformer-B~\citep{liu2021swin}, TimeSformer~\citep{bertasius2021space}, Video Swin Transformer-B~\citep{liu2022video}, and SlowFast~\citep{feichtenhofer2019slowfast} to represent content-aware, distortion-aware, and motion-aware features of videos. They incorporate these quality representations as supplementary supervisory information to train a lightweight BVQA model in a knowledge manner. These studies demonstrate the potential for BVQA models to benefit from various computer vision tasks. In this paper, we further demonstrate that BVQA models can achieve better performance with quality-aware pre-trained features. Moreover, Zhou \textit{et al.}~\citep{zhou2025adaptive} introduce ASAL, which effectively handles continually evolving, non-stationary quality distributions in long 360-degree VR videos through adaptive score alignment and memory-based continual learning; although this setting differs from the short, single-scene social-media videos considered in our work, such adaptive mechanisms provide a promising direction for extending BVQA models like RQ-VQA toward more dynamic and evolving video environments.

\section{Proposed Model}
\label{sec:method}
\begin{figure*}[t]
    \centering
    \includegraphics[width=1\linewidth]{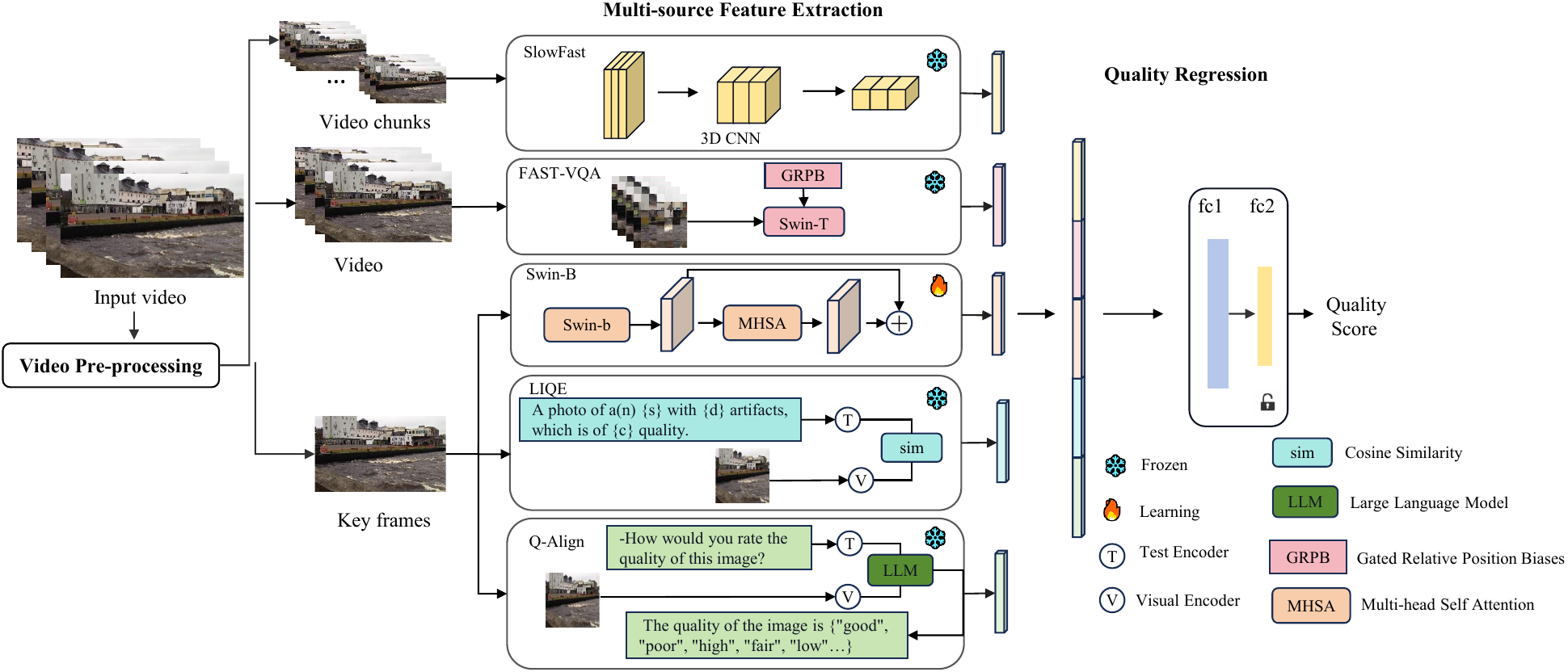}
    \caption{Overall architecture of the proposed \textbf{RQ-VQA} framework.  
Given an input video, key frames and chunks are first extracted for feature computation.  
The framework integrates multiple feature sources:  
(1) \textbf{Learnable spatial features} from Swin Transformer-B with an MHSA module,  
(2) \textbf{Temporal features} from the fast pathway of SlowFast,  
(3) \textbf{Spatiotemporal quality-aware features} from FAST-VQA,  
(4) \textbf{Spatial quality-aware features} from LIQE using cosine similarity between text and image embeddings, and  
(5) \textbf{Spatial quality-aware features} from Q-Align based on large multimodal model representations.  
All features are concatenated and fed into a two-layer MLP to regress the final quality score.  }
    \label{fig:model_framework}
\end{figure*}

As depicted in Figure~\ref{fig:model_framework}, our RQ-VQA framework comprises two main modules: a multi-source feature extraction module and a quality regression module. In the multi-source feature extraction module, we extract two types of features: (1) learnable features that capture quality-aware representations specific to the target video domain, and (2) off-the-shelf quality-aware features from existing quality assessment models, covering spatial, spatiotemporal, and temporal aspects. In the quality regression module, the extracted features are concatenated and passed through an MLP layer to predict the final quality score. The definitions of all variables and notations used throughout the framework are summarized in Table~\ref{tab:notation}.

\subsection{Video Pre-processing}
Given a video $\bm{x}= \{\bm x_i\}_{i=0}^{N-1}$, where $\bm x_i\in\mathbb{R}^{H\times W \times 3}$ represents the $i$-th frame. Here, $H$ and $W$ denote the height and the width of each frame respectively, and $N$ is the total number of frames. The features extracted by RQ-VQA can be categorized into three levels: spatial, temporal, and spatiotemporal. Therefore, we partition the video into three parts: key frames, video chunks, and the entire video. For key frames, we sample the first frame of every one-second video frame sequence as the key frame, denoted as:
\begin{equation}
\begin{aligned}
\bm z &= \{\bm z_i\}_{i=0}^{N_z-1}, \\
N_z &= N/r, \\
\bm z_i &= \bm x_{i*r}, \\
\end{aligned}
\end{equation}
where $r$ represents the frame rate of the video $\bm{x}$, $N_z$ is the total number of sampled key frames, and $\bm{z}_i$ is the $i$-th key frame. 

For video chunks, we split the video $\bm{x}$ into a series of video chunks:
\begin{equation}
\begin{aligned}
\mathcal{V} &= \{\bm v^{(i)}\}_{i=0}^{N_z-1}, \\
\bm v^{(i)} &= \{\bm x_s\}_{s=i*r}^{(i+1)*r-1}, \\
\end{aligned}
\end{equation}
where $\bm{v}^{(i)}$ denotes the $i$-th video chunk, containing $r$ consecutive frames, and each chunk corresponds to exactly one key frame.

For the entire video, the video $\bm{x}$ is directly used as the input.

\subsection{Multi-source Feature Extraction}
We extract quality-aware features from two sources: (1) learnable features from a base model fine-tuned on the target VQA datasets, and (2) off-the-shelf features from existing quality assessment models.

\subsubsection{The Base model}
Recent studies~\citep{sun2024analysis} indicate that most VQA datasets are predominantly affected by spatial distortions. Motivated by this observation, our base model is designed to focus on learning and extracting spatial features that are most relevant to visual quality perception in social media content. Towards this goal, we adopt Swin Transformer-B~\citep{liu2021swin}, a high-performance vision transformer architecture for its strong spatial modeling capabilities, as our backbone network. To enhance the model’s capacity to capture spatial regions that are most critical to perceptual quality, we incorporate a Multi-Head Self-Attention (MHSA) module~\citep{vaswani2017attention}, which enables the network to assign higher attention weights to salient areas that typically influence quality judgments in vide quality. Finally, global average pooling is applied to obtain the spatial quality representation. The overall process is defined as:
\begin{equation}
\begin{aligned}
\mathcal{F}^s_i &= {\rm GP(MHSA(SwinB}(\bm z_i))), \
\end{aligned}
\end{equation}
where $\rm GP$, $\rm MHSA$, and $\rm SwinB$ denote global average pooling, the MHSA module, and the Swin Transformer-B without its classification head, respectively. $\mathcal{F}^s_i$ represents the spatial features of the $i$-th key frame.

\subsubsection{Off-the-Shelf Quality Assessment Models}
We extract off-the-shelf features from three dimensions: spatial, spatiotemporal, and temporal, which serve as auxiliary features to quantify distortions along each corresponding dimension.

\vspace{0.2cm}
\noindent\textbf{Spatial Features:}  
We extract spatial features using two vision-language-based IQA models, LIQE~\citep{zhang2023blind} and Q-Align~\citep{wu2023q}, both trained on multiple quality assessment datasets to achieve strong generalization.

\textit{LIQE.} LIQE is a multitask-learning BIQA model built on CLIP~\citep{radford2021learning}, which consists of an image encoder and a text encoder. It employs a unified textual template  
\begin{equation}
\bm{t}_{\rm LIQE} = \text{``a photo of a(n) \{s\} with \{d\} artifacts, which is of \{c\} quality''}
\end{equation}
to jointly describe scene type $s$, distortion type $d$, and quality level $c$. Here, we consider nine scene categories: $s \in S =$ \{“animal”, “cityscape”, “human”, “indoor scene”, “landscape”, “night scene”, “plant”, “still-life”, and “others”\}, eleven distortion types: $d \in D =$ \{“blur”, “color-related”, “contrast”, “JPEG compression”, “JPEG2000 compression”, “noise”, “overexposure”, “quantization”, “under-exposure”, “spatially-localized”, and “others”\}, and five quality levels: $c \in C = \{1, 2, 3, 4, 5\} =$ \{“bad”, “poor”, “fair”, “good”, “perfect”\}. So, in total, we have $495$ text prompt candidates to compute the probabilities:
\begin{equation}
\begin{aligned}
\mathcal{F}^{\rm LIQE}_i &= {\rm LIQE} (\bm {z_i}, \bm{t}_{\rm LIQE}), \\
\end{aligned}
\end{equation}
where $\mathcal{F}^{\rm LIQE}_i$ represents the LIQE features of $i$-th key frames, which comprises $495$ dimensions corresponding to the scene category, artifact type, and quality level characteristics.

\textit{Q-Align.} Q-Align~\citep{wu2024q} is a large multimodal model designed for visual scoring, trained via instruction tuning on multiple IQA and VQA datasets. During training, MOS labels are converted into discrete text-defined rating levels (\textit{excellent, good, fair, poor, bad}) following ITU recommendations, and integrated into question–answer pairs for fine-tuning. The text input for Q-Align follows the format:  
\begin{equation}
\begin{aligned}
\bm{t}_{\rm{Q\text{-}Align}} &= \text{``How is the quality of this image? $|$img$|$} \\
&\quad \text{The quality of the image is [SCORE\_TOKEN]''},
\end{aligned}
\end{equation}
where \texttt{[SCORE\_TOKEN]} is a placeholder for the predicted quality level token.  

While Q-Align can output the quality rating token directly, we instead extract a richer representation by taking the feature map from its last hidden layer, followed by global average pooling:  
\begin{equation}
\mathcal{F}_i^{\mathrm{Q\text{-}Align}} = \mathrm{GP}(\mathrm{Q\text{-}Align}(\bm{z}_i, \bm{t}_{\rm{Q\text{-}Align}})),
\end{equation}  
where $\mathcal{F}_i^{\mathrm{Q\text{-}Align}} \in \mathbb{R}^{4096}$ denotes the Q-Align spatial features of the $i$-th key frame, and $\mathrm{Q\text{-}Align}(\cdot)$ represents the hidden-layer feature extraction.  

\vspace{0.2cm}
\noindent\textbf{Spatiotemporal Features:}  
We extract the spatiotemporal features using FAST-VQA~\citep{wu2022fast}, an efficient BVQA algorithm explicitly designed to model spatiotemporal quality representations while mitigating the high redundancy present in video data. To achieve this, it introduces the \textit{Grid Mini-Cube Sampling} (GMS) strategy, which pre-samples compact spatiotemporal \textit{fragments} from the video before feeding them into the backbone network, Video Swin Transformer Tiny~\citep{liu2022video}.  

Following the GMS process~\citep{wu2022fast}, each video frame is first divided into $G_f \times G_f$ uniform grids to ensure global coverage. From every grid, a raw-resolution patch is randomly sampled, and the sampling locations are kept temporally aligned across frames to preserve inter-frame quality variations. All sampled patches are then spatially reassembled in their original grid positions to form a set of spatiotemporal fragments, denoted as $\bm{x}^f$.

These spatiotemporal fragments are then fed into the Video Swin Transformer Tiny backbone to extract the quality-aware representation:  
\begin{equation}
\mathcal{F}^\mathrm{FAST\textrm{-}VQA} = \mathrm{FAST\textrm{-}VQA}(\bm{x}^f),
\end{equation}
where $\mathcal{F}^\mathrm{FAST\textrm{-}VQA}$ encodes both spatial structure and temporal motion cues.  

In our framework, we adopt FAST-VQA features pre-trained on the LSVQ~\citep{ying2021patch} dataset as the spatiotemporal quality representation.

\vspace{0.2cm}
\noindent\textbf{Temporal Features:}  
Temporal features capture motion-related information in videos, which is crucial for identifying distortions such as jitter from unstable camera operations or lag caused by low bandwidth during streaming. Following~\citep{sun2022deep, sun2024analysis}, we employ the fast pathway of SlowFast to extract motion features for each video chunk. The classification head of SlowFast is removed, and the temporal features are obtained by applying global average pooling to the last-stage feature maps:
\begin{equation}
\begin{aligned}
\mathcal{F}^t_i &= \mathrm{GP}(\mathrm{SlowFast}(\bm{v}^{(i)})),
\end{aligned}
\end{equation}
where $\mathrm{SlowFast}$ denotes the SlowFast backbone without the classification head, and $\mathcal{F}^t_i$ represents the temporal feature vector of the $i$-th video chunk.

\subsection{Quality Regression}

After extracting the multi-source features, we have:  
(1) \textit{Learnable features}, obtained from the based model that trained on target dataset, including the spatial features $\mathcal{F}^s_i$;  
(2) \textit{Off-the-shelf features}, derived from pre-trained quality assessment models, including the temporal features $\mathcal{F}^t_i$, the spatiotemporal features $\mathcal{F}^{\mathrm{FAST\textrm{-}VQA}}_i$, and the spatial quality-aware features $\mathcal{F}^{\mathrm{LIQE}}_i$ and $\mathcal{F}^{\mathrm{Q\textrm{-}Align}}_i$.   

We concatenate all features into the final representation $\mathcal{F}_i$:  
\begin{equation}
\begin{aligned}
&\mathcal{F}_i = \mathrm{Cat}(\mathcal{F}^s_i, \mathcal{F}^t_i, \mathcal{F}^{\mathrm{LIQE}}_i, \mathcal{F}^{\mathrm{Q\textrm{-}Align}}_i, \mathcal{F}^{\mathrm{FAST\textrm{-}VQA}}_i), \\
&\mathcal{F}^{\mathrm{FAST\textrm{-}VQA}}_i = \mathcal{F}^{\mathrm{FAST\textrm{-}VQA}},
\end{aligned}
\end{equation}
where $\mathrm{Cat}$ denotes the concatenation operator.  

A two-layer MLP network is then used to regress $\mathcal{F}_i$ into local quality scores $\hat{q}_i$:  
\begin{equation}
\begin{aligned}
\hat{q}_i &= \mathrm{MLP}(\mathcal{F}_i),
\end{aligned}
\end{equation}
where $\mathrm{MLP}$ denotes the multi-layer perceptron and $\hat{q}_i$ is the quality score of the $i$-th frame/chunk. The global quality score $\hat{q}$ is finally obtained by average pooling:  
\begin{equation}
\begin{aligned}
\hat{q} &= \frac{1}{N_z} \sum_{i=0}^{N_z-1} \hat{q}_i.
\end{aligned}
\end{equation}

\begin{algorithm}[htbp]
\caption{{RQ-VQA: Rich Quality-aware Feature Enabled VQA}}
\label{alg:rqvqa}
\small
\begin{algorithmic}[1]
\Require Video $\bm{x}=\{\bm{x}_i\}_{i=0}^{N-1}$ with frame rate $r$
\Ensure Predicted video quality score $\hat{q}$

\State \textbf{1. Pre-processing}
\State $N_z \gets \lfloor N/r \rfloor$
\State $\bm{z}_i \gets \bm{x}_{\,i\cdot r}$ \textbf{ for } $i=0,\ldots,N_z-1$ \Comment{key frames (1 fps)}
\State $\bm{v}^{(i)} \gets \{\bm{x}_s\}_{s=i r}^{(i+1)r-1}$ \textbf{ for } $i=0,\ldots,N_z-1$ \Comment{video chunks}
\Statex

\State \textbf{2. Learnable Spatial Features (Base Model)}
\For{$i=0$ to $N_z-1$}
  \State $\mathcal{F}^{s}_i \gets \mathrm{GP}\!\big(\mathrm{MHSA}(\mathrm{SwinB}(\bm{z}_i))\big)$
\EndFor
\Statex

\State \textbf{3. Off-the-Shelf Spatial Quality Features}
\For{$i=0$ to $N_z-1$}
  \State $\mathcal{F}_i^{\mathrm{LIQE}} \gets \mathrm{LIQE}(\bm{z}_i,\,\bm{t}_{\mathrm{LIQE}})$
  \State $\mathcal{F}_i^{\mathrm{Q\text{-}Align}} \gets \mathrm{GP}\!\big(\mathrm{Q\text{-}Align}(\bm{z}_i,\,\bm{t}_{\mathrm{Q\text{-}Align}})\big)$
\EndFor
\Statex

\State \textbf{4. Off-the-Shelf Spatiotemporal Quality Features}
\State $\bm{x}^{f} \gets \mathrm{GMS}(\bm{x})$
\State $\mathcal{F}^{\mathrm{FAST\textrm{-}VQA}} \gets \mathrm{FAST\textrm{-}VQA}(\bm{x}^{f})$
\Statex

\State \textbf{5. Off-the-Shelf Temporal Motion Features}
\For{$i=0$ to $N_z-1$}
  \State $\mathcal{F}^{t}_i \gets \mathrm{GP}\!\big(\mathrm{SlowFast}(\bm{v}^{(i)})\big)$
\EndFor
\Statex

\State \textbf{6. Multi-source Feature Fusion}
\For{$i=0$ to $N_z-1$}
  \State $\mathcal{F}_i \gets \mathrm{Cat}\!\big(\mathcal{F}^{s}_i,\,\mathcal{F}^{t}_i,\,\mathcal{F}^{\mathrm{LIQE}}_i,\,\mathcal{F}^{\mathrm{Q\text{-}Align}}_i,\,\mathcal{F}^{\mathrm{FAST\textrm{-}VQA}}\big)$
\EndFor
\Statex

\State \textbf{7. Quality Regression}
\For{$i=0$ to $N_z-1$}
  \State $\hat{q}_i \gets \mathrm{MLP}(\mathcal{F}_i)$
\EndFor
\State $\hat{q} \gets \frac{1}{N_z}\sum_{i=0}^{N_z-1}\hat{q}_i$
\Statex

\State \textbf{8. Optimization (training only)}
\State $\mathcal{L} \gets 1-\mathrm{PLCC}(q,\,\hat{q})$ \Comment{minimize over model parameters}
\State \Return $\hat{q}$
\end{algorithmic}
\end{algorithm}

\subsection{Loss Function}

We use the PLCC loss to optimize the proposed BVQA model:
\begin{equation}
\begin{aligned}
\mathcal{L} &= (1-\frac{\langle \bm{\hat{q}} - {\rm mean}(\bm{\hat{q}}), \bm {q} - {\rm mean}(\bm{q}) \rangle}{\Vert \bm{\hat{q}} - {\rm mean}(\bm{\hat{q}}) \Vert_2 \Vert {\bm q} - {\rm mean}(\bm{q}) \Vert_2})/2, \\
\end{aligned}
\end{equation}
where ${\bm q}$ and $\bm {\hat{q}}$ are the vectors of ground-truth and predicted quality scores of the videos in a batch respectively, $ {\langle \cdot \rangle}$ represents the inner product of two vectors, $\Vert\cdot\Vert$ denotes the norm operator for a vector, and $\rm mean$ is the average operator for a vector.

To provide a clear overview of how the aforementioned modules operate together, the full workflow of RQ-VQA is summarized in Algorithm~\ref{alg:rqvqa}.

\section{Experiment}
\subsection{Experimental Protocol}
\noindent\textbf{Test Datasets.}  
We evaluate our model on three VQA datasets—KVQ~\citep{lu2024kvq}, TaoLive~\citep{zhang2023md}, and LIVE-WC~\citep{yu2021predicting}—all of which target the quality assessment of social media videos.  
For KVQ, we train the model using the publicly available data from the NTIRE 2024 Short-form UGC Video Quality Assessment Challenge\footnote{\url{https://codalab.lisn.upsaclay.fr/competitions/17638}}, and evaluate it on both the validation and test sets.  
For TaoLive and LIVE-WC, we perform random scene-based splits of the videos with an $80\%$–$20\%$ train–test ratio, repeat the process five times, and report the average results.  
A detailed overview of these datasets is provided in Table~\ref{overview_vqa_dataset}.

\vspace{0.2cm}
\noindent\textbf{Implementation Details.} As stated in Section \ref{sec:method}, we utilize Swin Transformer-B~\citep{liu2021swin} and SlowFast R50~\citep{feichtenhofer2019slowfast} as the backbones of the spatial and temporal quality analyzers in the basic model. To improve the generalization ability of the basic model, we first train it on the LSVQ dataset~\citep{ying2021patch}, following the training strategy in~\citep{sun2024analysis}. Regarding the spatial quality analyzer, we resize the resolution of the minimum dimension of key frames as $384$ while preserving their aspect ratios. During the training and test stages, the key frames are randomly and centrally cropped with a resolution of 384$\times$384. As for the temporal quality analyzer, the resolution of the video chunks is resized to 224$\times$224 without respecting the aspect ratio. For LIQE, Q-Align, and FAST-VQA, we adhere to the original setups of these methods without making any alterations to extract the corresponding features. The Adam optimizer with the initial learning rate $1\times10^{-5}$ and batch size $6$ is used to train the proposed model on a server with $2$ NVIDIA RTX 3090. We decay the learning rate by a factor of $10$ after $10$ epochs and the total number of epochs is set as $30$.

\vspace{0.2cm}
\noindent\textbf{Compared Models.} We compare the proposed method with eight typical BVQA methods, including four knowledge-driven methods: NIQE~\citep{mittal2012making}, TLVQM~\citep{korhonen2019two}, VIDEVAL~\citep{tu2021ugc}, and RAPIQUE~\citep{tu2021rapique}, and four data-driven methods: VSFA~\citep{li2019quality}, SimpleVQA~\citep{sun2022deep}, FAST-VQA~\citep{wu2023exploring}, and Q-Align~\citep{wu2023q}. Except for Q-Align, we train other BVQA models for fair comparison.

\vspace{0.2cm}
\noindent\textbf{Evaluation Criteria.}
We employ two criteria to evaluate the performance of VQA models: PLCC and Spearman rank-order correlation coefficient (SRCC). Note that PLCC assesses the prediction linearity of the VQA model, while SRCC evaluates the prediction monotonicity. An outstanding VQA model should achieve SRCC and PLCC values close to 1.
Before computing PLCC, we adhere to the procedure outlined in~\citep{antkowiak2000final} to map model predictions to  MOSs by a monotonic four-parameter logistic function to compensate for prediction nonlinearity.

\begin{table*}
\small
\centering
\renewcommand{\arraystretch}{1.2}
\caption{Performance of the compared models and the proposed model on KVQ validation, KVQA test, TaoLive, and LIVE-WC datasets. The best-performing model is highlighted in each column}
\label{performance}
\resizebox{1\textwidth}{!}{
\begin{tabular}{cl|cc cc cc cc}
\toprule[.15em]
\multicolumn{2}{c|}{\multirow{2}{*}{BVQA Methods}} & \multicolumn{2}{c}{KVQ Validation} & \multicolumn{2}{c}{KVQ Test} & \multicolumn{2}{c}{TaoLive} & \multicolumn{2}{c}{LIVE-WC} \\
 &  & SRCC & PLCC & SRCC & PLCC & SRCC & PLCC & SRCC & PLCC \\
\hline
\multirow{6}{*}{\makecell[c]{Knowledge-driven \\Methods}}&NIQE~\citep{mittal2012making} &0.239 &0.241 & 0.272 & 0.281 & 0.331& 0.327& 0.245 & 0.241 \\
  &BRISQUE~\citep{mittal2012no} &0.472 &0.480 & 0.489 &0.493  &0.764 & 0.767& 0.794 &  0.797 \\
&TLVQM~\citep{korhonen2019two}  &0.490 &0.509 &  0.511& 0.524 & 0.869 &0.873 & 0.827 & 0.831 \\

&VIDEAL~\citep{tu2021ugc} & 0.369& 0.639&  0.425&0.652  &0.889 &0.892 &0.822  & 0.820  \\
&RAPIQUE~\citep{tu2021rapique} &0.803 & 0.801& 0.815 & 0.818 & 0.841 & 0.838 & 0.867 & 0.866 \\
\hline
\multirow{6}{*}{\makecell[c]{Data-driven \\Methods}}&VSFA~\citep{li2019quality}  & 0.830 & 0.834& 0.843 & 0.840 &0.904 & 0.903& 0.857 & 0.857 \\
& SimpleVQA~\citep{sun2022deep}&0.874 &0.875 & 0.881 & 0.877 & \textbf{0.916} &0.915 & 0.913 &  0.920\\
&FAST-VQA~\citep{wu2023exploring} &0.864 &0.865 &0.871  &  0.870 & 0.876 & 0.881 & 0.849 & 0.852 \\
&Q-Align~\citep{wu2023q} &0.703 &0.701 & 0.664 & 0.693 & 0.742& 0.722& 0.739 &  0.714 \\
& \textbf{RQ-VQA} & \textbf{0.914} & \textbf{0.918} & \textbf{0.926} & \textbf{0.924} & 0.912& \textbf{0.918}& \textbf{0.955} & \textbf{0.955}  \\
\bottomrule[.15em]
\end{tabular}
}
\end{table*}

\subsection{Experimental Results}

We present the experimental results in Table~\ref{performance}, from which several observations can be made.  
First, all knowledge-driven methods perform noticeably worse on the three social media VQA datasets, indicating that they struggle to capture the complex distortions and content characteristics present in social media videos. This limitation is likely due to their reliance on handcrafted features, which are less effective in modeling diverse and content-dependent degradations.  

Second, the proposed model consistently achieves the best performance on both the KVQ and LIVE-WC datasets, outperforming all competing BVQA methods by a clear margin in terms of both SRCC and PLCC. This highlights the effectiveness of our multi-source feature integration strategy, where the combination of learnable and off-the-shelf quality-aware features yields a richer and more discriminative representation, enabling more accurate quality prediction in complex BVQA scenarios (\textit{e.g.}, social media videos).  

Third, while the proposed model attains the highest average performance across all datasets, its improvement over SimpleVQA on TaoLive is relatively small, with the two methods showing comparable results. A possible explanation is that videos in TaoLive predominantly contain front-facing human subjects with relatively static backgrounds, and the distortion type is limited to compression artifacts. This setting is less challenging compared to KVQ and LIVE-WC, reducing the advantage of incorporating additional quality-aware features.

\begin{table}
\scriptsize
\centering
\renewcommand{\arraystretch}{1.2}
\caption{The results of the NTIRE 2024 Short-form UGC Video Quality Assessment Challenge~\citep{li2024ntire}}
\label{challenge}
\begin{tabular}{cc }
\toprule[.15em]
Team  & Final Score \\
\hline
SJTU MMLab (Proposed) & 0.9228 \\
IH-VQA & 0.9145 \\
TVQE & 0.9120 \\
BDVQAGroup & 0.9116 \\
VideoFusion & 0.8932 \\
MC$^2$Lab & 0.8855 \\
Padding & 0.8690 \\
ysy0129 & 0.8655 \\
lizhibo & 0.8641 \\
YongWu & 0.8323 \\
we are a team & 0.8243 \\
dulan & 0.8098 \\
D-H & 0.7677 \\
\bottomrule[.15em]
\end{tabular}
\end{table}

\subsection{Experiment Results on NTIRE Challenge}

The proposed RQ-VQA method participated in the NTIRE 2024 Short-form UGC Video Quality Assessment Challenge~\citep{li2024ntire}, which was organized as part of the NTIRE 2024 Workshop at CVPR 2024. This international competition was jointly hosted by the Computer Vision Lab of ETH Zürich and Kuaishou (Kwai) Technology. The challenge aimed to advance research on short-form user-generated content (S-UGC) video quality assessment, leveraging the newly released KVQ dataset~\citep{lu2024kvq}, which includes 4,200 short videos collected from the Kwai platform. These videos span nine content categories and multiple processing workflows such as enhancement, pre-processing, and transcoding. The dataset was divided into training, validation, and testing subsets with a ratio of 7:1:2.

A total of over 200 participants registered for the competition, among which 49 teams submitted results during the development phase, and 13 finalist teams provided complete fact sheets and codes for the final evaluation. The performance of all teams was assessed using four quantitative metrics—SROCC, PLCC, Rank1, and Rank2—which jointly measure the correlation and ranking consistency of predicted quality scores. The final score was calculated as:
\begin{equation}
\text{Score} = 0.45 \times \text{SRCC} + 0.45 \times \text{PLCC} + 0.05 \times \text{Rank1} + 0.05 \times \text{Rank2}.
\end{equation}
This composite metric jointly evaluates both correlation accuracy and fine-grained ranking consistency, ensuring a balanced assessment of model performance in the competition.
 
To enhance robustness, we performed ten random $80\%$–$20\%$ splits of the public training set from the KVQ dataset, trained models on each split, and ensembled their predictions to obtain the final performance.  
As shown in Table~\ref{challenge}, RQ-VQA (SJTU MMLab) achieved a score of $0.9228$, ranking first among all participating teams and surpassing the second-best method by a considerable margin.  
These results further demonstrate the effectiveness and generalization ability of our multi-source feature integration strategy in real-world competition settings.

\subsection{Ablation Studies}
\begin{table}[t]
\scriptsize
\centering
\renewcommand{\arraystretch}{1.2}
\caption{Ablation study on the quality-aware features using the KVQ test set. ``Spatial'' and ``Spatiotemporal'' indicate the feature type.}
\label{abliation_experiment}
\begin{tabular}{cccc|cc}
\toprule[.15em]
\multirow{2}{*}{\makecell[c]{Base Model \\ + SlowFast}} & 
\multirow{2}{*}{\makecell[c]{Q-Align \\ (Spatial)}} & 
\multirow{2}{*}{\makecell[c]{LIQE \\ (Spatial)}} & 
\multirow{2}{*}{\makecell[c]{FAST-VQA \\ (Spatiotemporal)}} & 
\multicolumn{2}{c}{KVQ Test} \\
 & & & & SRCC & PLCC \\
\hline
$\surd$ & $\times$ & $\surd$ & $\surd$ & 0.922 & 0.920 \\
$\surd$ & $\surd$ & $\times$ & $\surd$ & 0.923 & 0.921 \\
$\surd$ & $\surd$ & $\surd$ & $\times$ & 0.924 & 0.925 \\
$\surd$ & $\surd$ & $\surd$ & $\surd$ & 0.926 & 0.924 \\
\bottomrule[.15em]
\end{tabular}
\end{table}

\begin{table*}[t]
\centering
\scriptsize
\renewcommand{\arraystretch}{1.2}

\begin{minipage}{0.48\textwidth}
\centering
\caption{Ablation study on different loss functions for RQ-VQA using the KVQ validation set}
\label{abliation_experiment_loss_function}
\begin{tabular}{c|cc}
\toprule[.15em]
\multirow{2}{*}{\makecell[c]{Loss Function}} & 
\multicolumn{2}{c}{KVQ Validation} \\
 &  SRCC & PLCC \\
\hline
MSE &  0.897 & 0.906 \\
MAE  &  0.907 & 0.912 \\
PLCC  & 0.914  &  0.918\\
\bottomrule[.15em]
\end{tabular}
\end{minipage}
\hfill
\begin{minipage}{0.48\textwidth}
\centering
\caption{Ablation study on different input resolutions for RQ-VQA using the KVQ validation set}
\label{abliation_experiment_resolution}
\begin{tabular}{c|cc}
\toprule[.15em]
\multirow{2}{*}{\makecell[c]{Input Resolution}} & 
\multicolumn{2}{c}{KVQ Validation} \\
 &  SRCC & PLCC \\
\hline
$224\times224$ &  0.886 & 0.901 \\
$320\times320$  &  0.892 & 0.907 \\
$380\times380$  &  0.914 & 0.918 \\
\bottomrule[.15em]
\end{tabular}
\end{minipage}

\end{table*}

\noindent\textbf{Network Structure.} We individually remove Q-Align (spatial), LIQE (spatial), and FAST-VQA (spatiotemporal) from the full model and evaluate the resulting variants on the KVQ test set. 
\textbf{Note} that the temporal features extracted by SlowFast are \emph{not} part of this ablation, because SlowFast is pretrained for \emph{action recognition} rather than quality assessment; our goal here is to isolate the effect of features explicitly designed for quality assessment. 
The results in Table~\ref{abliation_experiment} show that removing any quality-aware feature leads to a drop in both SRCC and PLCC, indicating that each source provides complementary information for prediction. 
When all three quality-aware features are integrated with the base model, the proposed method achieves the best performance, confirming the effectiveness and synergy of the extracted features.

\vspace{0.2cm}
\noindent\textbf{Input Resolution.} We evaluate the effect of input resolution by testing $224\times224$, $320\times320$, and $380\times380$ settings, as summarized in Table~\ref{abliation_experiment_resolution}. 
The results demonstrate a steady improvement in both SRCC and PLCC with higher spatial resolution, reaching the best performance at $380\times380$. 
This indicates that richer spatial details help the model capture fine-grained perceptual cues, leading to more stable and accurate predictions within the RQ-VQA framework.

\vspace{0.2cm}
\noindent\textbf{Loss Function.} To further analyze the optimization objective, we compare different loss functions, including Mean Squared Error (MSE), Mean Absolute Error (MAE), and PLCC losses, on the KVQ validation set. 
As reported in Table~\ref{abliation_experiment_loss_function}, the PLCC loss achieves the highest correlations, suggesting that directly optimizing for correlation yields more consistent alignment with subjective MOSs. 
This result validates the rationality of employing a correlation-oriented objective in RQ-VQA.

\begin{table}[t]
% \scriptsize
\renewcommand{\arraystretch}{1.2}
\footnotesize
\centering
\caption{Computational complexity of the proposed RQ-VQA framework. All measurements were conducted on an NVIDIA RTX 3090 GPU. FLOPs and inference time are computed for an eight-second 1080p video}
\label{tab:comptatial_complexity}
\resizebox{1\textwidth}{!}{
\begin{tabular}{l|ccccc|cc}
\hline
Component & \textbf{Base Model} & \textbf{SlowFast} &\textbf{ LIQE} & \textbf{Q-Align} & \textbf{FAST-VQA} & \textbf{Trained} & \textbf{Overall} \\ 
\hline
Parameters (M) & 86.83 & 33.64 & 59 & $8.2\times10^3$ & 27 & 86.83 & $8.4\times10^3$ \\ 
FLOPs (G) & 357.62 & 50.58 & 707.2 & $12\times10^3$ & 279 & --- & $13.34\times10^3$ \\ 
Inference Time (s) & 0.05 & 0.19 & 0.76 & 4.20 & 0.11 & --- & 5.30 \\ 
\hline
\end{tabular}
}
\end{table}

\subsection{Computational Complexity}

\begin{figure*}[t]
    \centering
    \includegraphics[width=1\linewidth]{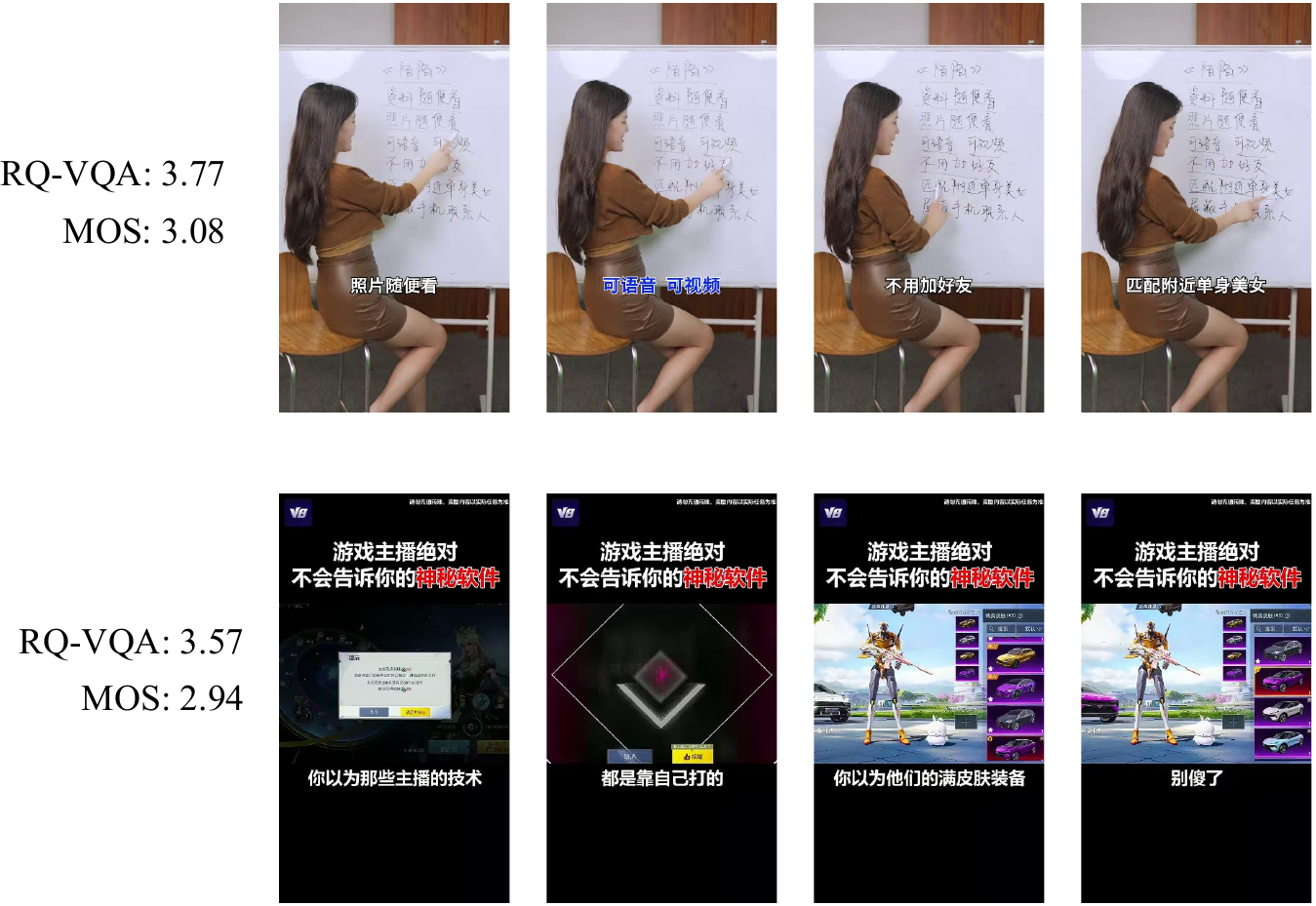}
    \caption{Typical failure cases of RQ-VQA on social-media videos. In both examples, the predicted RQ-VQA scores are noticeably higher than the corresponding MOS values, indicating overestimation on text-overlay or synthetic-content scenes.}
    \label{fig:video_frames_comparison}
\end{figure*}

To assess the practicality and deployment efficiency of the proposed RQ-VQA framework, we provide a comprehensive analysis of its computational costs, including the number of  parameters, floating-point operations (FLOPs), and inference time. We report these metrics for each major component—Base Model, SlowFast, LIQE, Q-Align, FAST-VQA—as well as for the complete system. All measurements were conducted using an NVIDIA RTX 3090 GPU, and the inference time corresponds to processing an eight-second 1080p video.

As summarized in Table~\ref{tab:comptatial_complexity}, the full RQ-VQA model contains approximately ${8.4\times10^3}$M parameters and requires ${13.3\times10^3}$G FLOPs, resulting in an average inference time of ${5.30}$ seconds. This demonstrates that the  proposed framework can operate at near real-time speed, making it suitable for large-scale quality monitoring and practical streaming-media applications. Among all modules, Q-Align  contributes the most to the overall model size due to its large multimodal backbone. However, this module is kept frozen during training and is only invoked once for feature extraction,  thus introducing no additional optimization overhead. The remaining components—including the base model, SlowFast, LIQE, and FAST-VQA—are comparatively lightweight and efficient. Furthermore, in resource-constrained scenarios, the Q-Align branch can be removed to  substantially reduce computational cost while still maintaining competitive performance, as  validated in Table~\ref{abliation_experiment}. This flexibility highlights the modular nature of RQ-VQA and its ability to adapt to different deployment constraints.

\subsection{Discussion}

Although RQ-VQA achieves state-of-the-art performance on multiple social-media video datasets, several limitations can still be observed. As illustrated in Figure~\ref{fig:video_frames_comparison}, our model tends to \textbf{overestimate the perceptual quality} of videos containing \textbf{text-overlaid human-subject scenes} or \textbf{synthetic gaming elements}. In both examples, the predicted RQ-VQA scores (3.77 and 3.57) are notably higher than the corresponding human opinion scores (MOS = 3.08 and 2.94). This overestimation primarily occurs because the pretrained quality-aware features employed in RQ-VQA were originally derived from \textbf{natural photographic content}, which exhibits smooth textures and organic distortions. In contrast, social-media videos often contain \textbf{text–image mixed compositions}, \textbf{graphic overlays}, or \textbf{stylized synthetic artifacts}, whose sharp edges and uniform color regions are sometimes misinterpreted by pretrained networks as indicators of clarity or high perceptual quality.

These findings suggest that, while the integration of spatial, temporal, and spatiotemporal cues provides robust generalization for natural content, it remains suboptimal for videos with \textbf{non-natural or artificial visual structures}. Future work will therefore explore extending RQ-VQA with \textbf{domain-specific feature encoders} and \textbf{text- and graphic-aware representations}, to enhance its reliability and perceptual alignment across diverse social-media video types.

\section{Conclusion}
In this paper, we presented \textbf{RQ-VQA}, a simple yet effective blind video quality assessment framework that leverages \emph{rich quality-aware features} from multiple off-the-shelf BIQA and BVQA models.  
By integrating domain-specific learnable features from a base model with spatial, spatiotemporal, and temporal quality-aware representations extracted from pretrained models, RQ-VQA effectively exploits the complementary strengths of different feature sources.  
Extensive experiments on three public social media VQA datasets demonstrated that our method achieves state-of-the-art performance, while maintaining a straightforward training pipeline.  
Moreover, RQ-VQA ranked first in the CVPR NTIRE 2024 Short-form UGC Video Quality Assessment Challenge, further validating its robustness and generalization ability.  
Given its simplicity and flexibility, the proposed framework can be readily extended to other quality assessment tasks by incorporating additional quality-aware feature sources.

% %% The Appendices part is started with the command \appendix;
% %% appendix sections are then done as normal sections
% \appendix
% \section{Example Appendix Section}
% \label{app1}

% Appendix text.

% %% For citations use: 
% %%       \citep{<label>} ==> [1]

% %%
% Example citation, See \citep{lamport94}.

%% If you have bib database file and want bibtex to generate the
%% bibitems, please use
%%
 \bibliographystyle{elsarticle-harv} 
 \bibliography{paper_references.bib}

%% else use the following coding to input the bibitems directly in the
%% TeX file.

%% Refer following link for more details about bibliography and citations.
%% https://en.wikibooks.org/wiki/LaTeX/Bibliography_Management

% \begin{thebibliography}{00}

% %% For numbered reference style
% %% \bibitem{label}
% %% Text of bibliographic item

% \bibitem{lamport94}
%   Leslie Lamport,
%   \textit{\LaTeX: a document preparation system},
%   Addison Wesley, Massachusetts,
%   2nd edition,
%   1994.

% \end{thebibliography}
\end{document}